\documentclass[twocolumn,pre]{revtex4} 
\usepackage{graphicx}
\usepackage{dcolumn}

\def\ssout{_{\rm out}}
\def\ssin{_{\rm in}}
\def\LP{\left(}
\def\RP{\right)}
\newcommand{\be}{\begin{equation}} 
\newcommand{\ee}{\end{equation}} 
\newcommand{\bd}{\begin{displaymath}} 
\newcommand{\ed}{\end{displaymath}} 
\newcommand{\bea}{\begin{eqnarray}} 
\newcommand{\eea}{\end{eqnarray}} 
\newcommand{\beay}{\begin{eqnarray*}} 
\newcommand{\eeay}{\end{eqnarray*}} 
\newcommand{\bc}{\begin{center}} 
\newcommand{\ec}{\end{center}}

\begin{document} 
\title{Random model for RNA interference yields scale free network}

\author{ Duygu Balcan$^{1}$ and Ay\c se Erzan$^{1,2}$ } 
\affiliation{$^1$ Department of Physics, Faculty of Sciences and 
Letters\\
Istanbul Technical University, Maslak 34469, Istanbul, Turkey \\
$^2$ G\"ursey Institute, P.O.B. 6, \c Cengelk\"oy, 34680 Istanbul, 
Turkey
}
\date{\today}

\begin{abstract}

We introduce a random bit-string model of  post-transcriptional genetic 
regulation based on 
sequence matching.  The model spontaneously yields a scale free network with 
power law scaling 
with $ \gamma=-1$ and also exhibits 
log-periodic behaviour.  The in-degree distribution is much narrower, and 
exhibits a pronounced peak followed by a Gaussian distribution.  The network is 
of the smallest world type, with the average  minimum path length independent of 
the size of the network, as long as the network consists of one giant cluster.  The percolation threshold  depends on the system size.


\end{abstract}
\maketitle

\section{Introduction}

Although biology on the whole is a ``knowledge based'' discipline, with a strong 
traditional bias 
towards a reverse engineering approach to evolution, and a  ``form follows 
function'' approach, 
prominant workers in the field have stressed that natural selection could very 
well have operated 
on complex structures already present in the pre-biotic world.  
Eigen~\cite{Eigen} pointed out 
that non-linear out of equilibrium systems were capable of amplifying random 
fluctuations 
participating in feed-back loops,
while Kauffman~\cite{Kauffman1,Kauffman2} introduced random Boolean 
networks as null models of the 
genetic regulatory mechanism, and showed that they could spontaneously give rise 
to a great 
degree of complexity.  Meanwhile, within the last  two decades, we have learned 
a great deal 
about Self-Organized Criticality~\cite{Bak}, namely the spontaneous emergence of 
scale free 
structures in open systems far from equilibrium, driven by conserved fluxes. 
Thus,
within a statistical mechanics context, it is much more natural to consider an 
ensemble of different states among which complex structures arise
spontaneously. Evolutionary processes can be 
regarded as inducing dynamics on the distribution of states in this high 
dimensional phase space.

In this paper we  introduce a null model for gene interactions resulting in the 
regulation of gene 
expression.  This model is based on sequence matching  on a random bit-string 
representation of 
the chromosome and 
is therefore radically different from the random Boolean networks which have 
been considered 
before~\cite{Kauffman1,Kauffman2,Weisbuch,Derrida1,Derrida2,Flyvberg}.  It can be considered as an out of equilibrium system subject to a constant mutation rate, achieving a steady state invariant under further mutations.

\begin{figure}
\bc
\leavevmode
\includegraphics[width=8.5cm,angle=0]{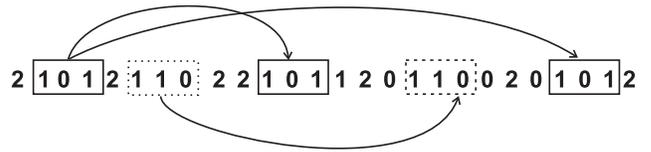}
\caption{ The random sequence of symbols representing a chromosome.  The ``2" 
represents a start or stop sign for a gene, and 
the 0's and 1's code the genes.  The arrows indicate that one gene sequence is 
embedded in another one.} \label{Chromosome}
\ec
\end{figure}

We present simulation results which display many qualitative features of
gene regulation networks found in nature~\cite{Maslov1,Maslov2}.  We find 
that the in- and out- degree distributions
are qualitatively different from each other, the out-degree distribution 
exhibiting power
law decay with $n(k\ssout) \sim k\ssout ^\gamma$, with $\gamma=-1$ and log-
periodic oscillations for relatively small $k$.  The in-degree
distribution, on the other hand, is much more localized.

The network has smallest world characteristics, with the
cluster diameter, or the average minimum path length being 
essentially independent of the 
cluster size as long as the network consists of one cluster.

The average 
clustering coefficients for the in and out bonds  have been calculated as 
0.648 and 0.034 respectively.  It is interesting to note that the 
clustering coefficients for the in-bonds behave much like for  classical 
random networks~\cite{Erdos}, while the out-bonds have a 
clustering coefficient typical of scale free 
networks.~\cite{Dorogovtsev,Barabasi}

%Kumelenme katsayilari soyle:
%<Cout>=0.034035
%<Cin>=0.648129
%<C>=0.534313  

In section 2, we will first motivate and then define the model.  In
section 3, we will present our simulation results.  In section 4, we
consider a toy model which mimics some of the features we observe in our
simulations, and indicate ways in which this toy model may be improved in
order to provide insights into how our basic model works.  Section 5
contains our conclusions and a discussion.

\section{Modelling RNA interference}
\subsection{Genomic regulatory networks}

Protein networks, which are an important component of transcriptional gene
regulation networks~\cite{biobook1,Sole}, display a scale free structure, 
with
the out-degree distribution characterised by a power law $n(k) \sim
k^{\gamma}$, with the exponent $\gamma \simeq - 2.5$.~\cite{Maslov1,Maslov2}
Gene regulation networks actually operate at many different
levels.~\cite{Maniatis} Post-transcriptional gene regulation, or RNA
interference~\cite{Hannon}, is a mechanism where RNA strips may go and
directly bind upon complementary segments on messenger RNA destined to be
translated into some protein, thereby suppressing the production
of this protein.  Although we are not aware of a scaling analysis of
post-transriptional gene interaction, it would be a fair guess to assume
that interaction complexes of various sizes may arise in this type of
interaction as well.

In all protein-protein or intra-genomic interactions, as well as the 
transcription and translation 
mechanism itself, essential lock-and key mechanisms are in
operation.  For normal translation to take place, rRNA in the ribosomes
must be able to recognize and match the different amino acids and the
corresponding three-letter anti-codon on the mRNA. In this, the rRNA is
aided by the intermediary tRNA, which assumes a very specific three
dimensional structure depending on the amino acid to which it binds.
In transcriptional gene regulation, certain proteins known as
transcription factors (TF)  must first be synthesized, and then go and
bind onto specific ``promoter'' sites preceding the coding part of a gene,
in order that the RNA polymerase might start transcribing the DNA code
into the mRNA.  Conversely, the binding of other proteins onto the same
promoter sites may block the binding of the TF, and thereby block the
production of the mRNA.~\cite{Hannon,biobook2} In both cases, such binding
presupposes steric and chemical specificity. In RNA interference, the
short interfering RNA (siRNA) strips bind onto complementary sequences on 
the
mRNA, via Watson-Crick base pairing.~\cite{Hannon,biobook3}

Although it seems, from the above, as if there is a great diversity in
these lock-and-key mechanisms, it should be realised that they all
eventually match linear codes, even though this matching may take a few
intermediary steps. The three dimensional structures (so called secondary
structures) which come into play either in tRNA or the TF, are actually
determined by either the sequence of ribo-nucleic acids on the tRNA, or the
sequence of amino acids (primary structure) of the protein constituting
the TF.  Clearly the simplest is direct Watson-Crick base pairing between
complementary sequences, and it is this latter, as it appears in RNA
interference, which we will take to be the paradigm for our model.  In
fact we will further simplify the matching condition to consist of the
identity relation rather than complementarity, since both are one-to-one,
we believe that this should not change our results.

\begin{figure}
\bc
\leavevmode
\includegraphics[width=8.5cm,angle=0]{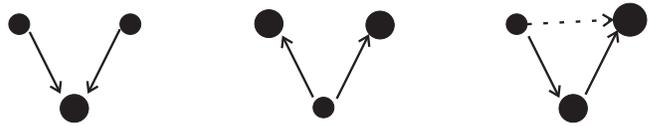}
\caption{ Different kinds of vertices allowed on our directed random
network. While the in or out-neighbors of a vertex may or may not be
connected to each other, in the case of the third configuration,
with a pair of mixed bonds, the neighboring nodes are necessarily
connected as shown, due to transitivity.} \label{vertices}
\ec
\end{figure}

\subsection{The Model}

The model is defined as follows.  We postulate a ``chromosome'' to consist of a 
sequence of fixed 
length L, of independently and identically distributed random numbers  with the 
probability distribution 
\be 
P(x) = p \delta (x-2) + (1-p)/2 [ \delta (x-1) + \delta (x)]
\ee

We define a ``gene'' to consist of a sequence 
of 0's and 1's situated between the $i$th and $i+1$st occurance of the symbol 
``2,'' and we will denote the $i$th gene by 
\be 
G_i= \{ x_{i,1}, x_{i,2},\ldots x_{i, \ell_i} \} \;\;\; i=1, \ldots, s
\ee
where $x_{i,\mu} \neq 2, \;\; \mu=1,\ldots \ell_i$ and $\ell_i$ is the 
length of the $i$th gene.
We have used periodic boundary conditions, but one could just as well agree to 
end the chromosome always with a ``2'' at the $L+1$st site.
Then
\be
\sum_i^s \ell_i = L-s\;\;\;,
\ee
where $s$ is the number of genes (the number of times the symbol 2 appears) on the chromosome.  Let  $n_\ell$ be the number 
of 
genes of length $\ell$.  It obeys the sum rule
\be
\sum_{\ell = 0}^{L-s}  n_\ell = s\;\;\;.
\ee
For a given number $s$ of genes, the number of possible realisations of a given 
set $\{ n_\ell \}$ is 
\be K[\{n_\ell \}] = { s! \over \prod_{\ell=0}^{L-s} n_\ell !} \;\;\;, \ee
and  the most probable distribution
${\overline n}(\ell)$ can easily be found by using  Lagrange multipliers, to be 
$\overline{n}(\ell)= L p^2 (1-p)^\ell$, in the limit of large $L$.

With these definitions we obtain a sequence of genes, seperated from each other 
by the 
symbol 2.  (See Fig. (\ref{Chromosome})). In case there are more than one 
consecutive 2's, they will be considered to bracket null genes.   Clearly, for 
large L, the number of non-null genes, $N$, will fluctuate around 
$\overline{N}= Lp-Lp^2$.

Each of the non-null genes constitutes a node in our gene 
regulation network.  The interactions do not depend on the proximity of the 
genes along 
the chromosome. We define the adjacency matrix $w_{ij}$ by the matching 
condition such 
that 
\be
w_{ij} = \cases{1 & $G_i\subset G_j$ \cr 0 & otherwise}\;\;\;.
\ee
By $G_i\subset G_j$ we mean  $x_{i,\mu} = x_{j,\mu+\nu}$ for 
$\mu=1,\ldots\ell_i$ for at least one integer  $\nu$ such that $0\leq \nu \leq 
\ell_j-\ell_i$. 
Note that 
$w_{ij}=1$ implies that $\ell_i\le \ell_j$; in the case of the equality, 
$w_{ij}=1$ if and only if  the two sequences $G_i$ and $G_j$ are 
identical, i.e., congruent.
Thus, two genes are said to  interact if 
the sequence $G_i$ occurs at least once as an unbroken subsequence of 
$G_j$, i.e.,
if one can be embedded in the other at least once.  

Clearly this adjacency (or  connectivity) matrix is directed.  Moreover 
connectivity is ``transitive''  in the sense that 
$w_{ij}=w_{jk}=1$ implies that $w_{ik}=1$.  The latter condition gives rise to a 
preferential 
attachment of incoming bonds to large genes, while small genes have an enhanced 
distribution of out-bonds.  However, we will see in the next section that the degree distribution is scale free for the out-bounds, but not for the in-bonds.

A simple argument tells us that our network is of the smallest 
world~\cite{Dorogovtsev} type.
If we take the out-bonds, we see that due to the transitivity, any two 
successive edges linking say vertex $ij$ and $jk$, necessarily imply the 
existence of another directed edge $ik$.  Thus $l^{(o)}_{\rm min} \equiv 1$, as 
long as the network consists of a single cluster. 
The in-bonds follow the same argument, giving 
$l^{(i)}_{\rm min}\equiv 1$ for a network with one giant cluster.

The question of whether a giant cluster always exists or whether we can identify the analog of a ``percolation threshold'' we will address in the next section, where we will also report numerical results for the minimum path length for undirected bonds (i.e., ignoring the directionality of the edges.)  We expect that the qualitative behaviour of the network should not depend on $p$ as long as $p$ is bounded away from the percolation threshold. 
Requiring the number of vertices, $N$,  to be larger than unity, 
i.e., 
$\overline{N} = Lp - L p^2 = Lp(1-p) > 1$ gives  $p(1-p)> 1/L$ (however this lower limit turns out not to be tight enough, i.e., $p_c>1/L$).
The average gene 
size, again in the limit of large $L$ is 
$\langle \ell \rangle = (1-p)/p$. The obvious requirement for a non-trivial 
network, that $\langle \ell \rangle > 1$ yields $p < 1/2$.
 We therefore expect 
to find scaling behaviour, if any, for $p_c < p < 1 / 2 $.

\begin{figure}
\bc
\leavevmode
\includegraphics[width=9cm,angle=0]{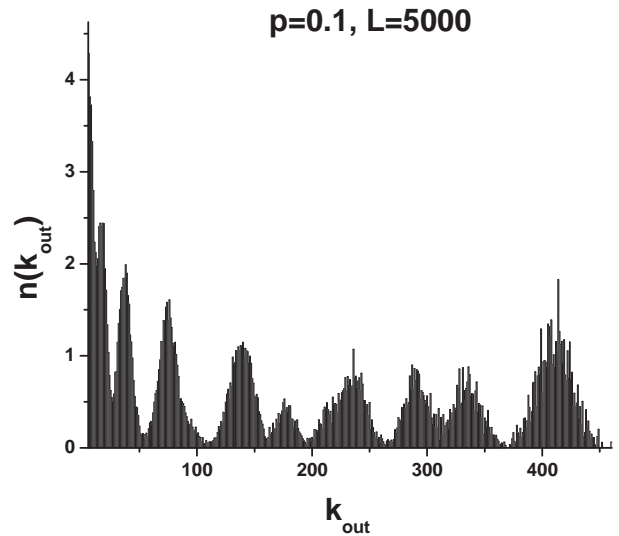}
\caption{The out degree distribution, for $p=0.1$ and $L=5000$. } \label{zout1}
\ec
\end{figure}

\begin{figure}
\bc
\leavevmode
\includegraphics[width=8.5cm,angle=0]{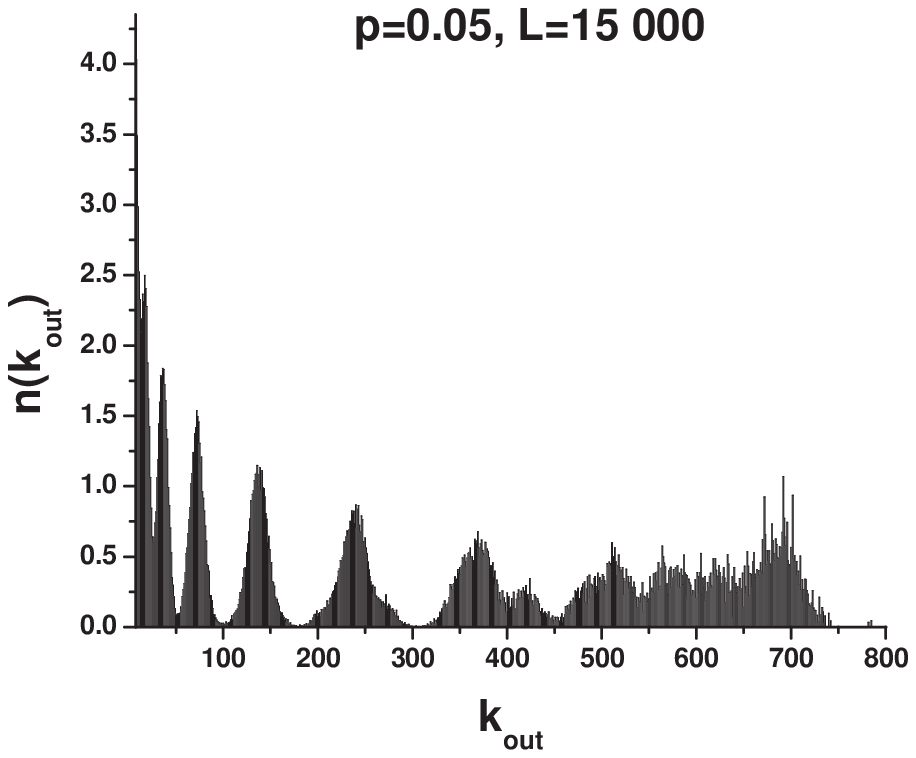}
\caption{The out degree distribution for $p=0.05$ and $L=15000$. } \label{zout2}
\ec
\end{figure}

The elements $w_{ij}$ of the connectivity matrix are equal to unity with
probabilities 
\be P_{ij}=P(\ell_i, \ell_j)\;\;, \label{P}
\ee
 which depend on the lengths
$(\ell_i, \ell_j)$ through $\ell_i$ and $\nu=\ell_j-\ell_i$.  It is
trivial to see that
\be
P(\ell, \ell) =   \LP {1 \over 2} \RP^{\ell} \;\;\;.
\ee

We may make a mean field theory type of approximation to  $P(\ell, 
\ell+\nu)$ by neglecting the correlations between overlapping subsequences of 
$G_j$, and obtain,
\be P(\ell, \ell+\nu) \simeq \LP {1 \over 2}\RP^{\ell} (1+\nu)\;\;\;.
\ee
In the Appendix we have computed $P(\ell, \ell+\nu)$ explicitly for $\nu = 
1,2$. However, so far it has not been possible  to extract the form of the 
degree distribution analytically.

The directed graph gives rise to different kinds of vertices, shown in 
Fig.(\ref{vertices}), which allow different classes of clustering coefficients 
$C$.  Let $k_{\rm out}(i)$, $k_{\rm in}(i)$, $k(i)$, be, respectively, the out-
degree, in-degree and the total degree of the vertex $i$. The clustering 
coefficient at a given vertex $i$ is defined as 
\be
C_i = {2 E(i) \over k(i) [k(i)-1]}
\ee
where $E(i)$ is the number of edges connecting the nearest neighbors of 
$i$.  
Thus it is the number of pairs of nearest neighbors directly connected to each 
other, normalized by the largest number of such connections possible.  We may 
extend this concept to directed graphs and define,
\be C\ssout (i) = {2 E\ssout(i) \over k\ssout (i) [k\ssout(i) -1]}
\ee
 and similiarly for $C\ssin$, where $ E\ssout (i)$ (respectively $E\ssin 
(i)$) 
are the number of edges connecting out (in) nearest neighbors of $i$.  Note that 
any pair of incoming and outgoing bonds passing through $i$ necessarily defines 
a triangle, due to transitivity, as shown in Fig.(~\ref{vertices}).  Thus, the clustering coefficient may be 
conveniently decomposed as 
\be C_i = { k\ssin(i)\, k\ssout(i) + E\ssin(i) + E\ssout(i) \over k (i) 
[k(i)-1]/2 } 
\;\;\;.
\ee

This is a null model in the sense that no assumptions have been made as to the 
fitness of any 
particular type of interaction; the resulting interactions depend only on the 
random sequences 
coded in the genes and on the distribution of gene lengths.  This random network 
provides the ``tabula rasa'' on which 
we assume natural selection will subsequently act.  As such, it is of great 
interest to determine the 
properties of the null network, which turn out to be highly non-trivial.  This 
is the task to which we turn in the next section.

\section{Simulation results}

To characterize the network defined by our model, we generated random chromosomes as defined above. The statistical properties of the network obtained from the totally random chromosome were checked to be invariant under a constant mutational load, with a mutation probability of 0.01. A mutation is affected as follows.
If the symbol ($x$) occupying the site to be mutated 
happens to be a 0 or 1, then it is flipped, i.e., we
set $x={\rm mod}_2(x+1)$.  If $x=2$, then it exchanges places with either
its right or left nearest neighbor, with equal probability.  It should be
noted that the first case corresponds to a substitutional mutation,
whereas the second to a shifting of the position of the start sign which
shifts the reading frame of the gene.  The latter gives rise to
far-reaching modifications, since the three-letter codes corresponding to
the different amino acids will be completely modified along the whole gene
if the reading frame is shifted.

The most remarkable property of the random network generated in this way
is that it has a scale free out-degree distribution, and a qualitatively
different in-degree distribution which is much less broad, with a
pronounced narrow peak followed by a Gaussian peak.  These qualitative
behaviours match the results found on protein and genomic networks to a
surprising extent.~\cite{Maslov1,Maslov2}

\begin{figure}
\bc
\leavevmode
\includegraphics[width=8.5cm,angle=0]{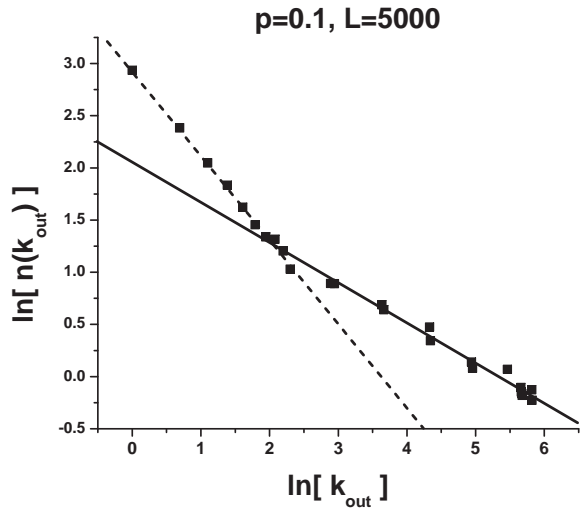}
\caption{The log-log fit to the envelope of the out-degree distribution
shown in Fig.(\ref{zout1}).  The slopes of the dashed and continuous lines 
are -0.9 and -0.39, respectively. } \label{fits1}
\ec
\end{figure}

\begin{figure}
\bc
\leavevmode
\includegraphics[width=8.5cm,angle=0]{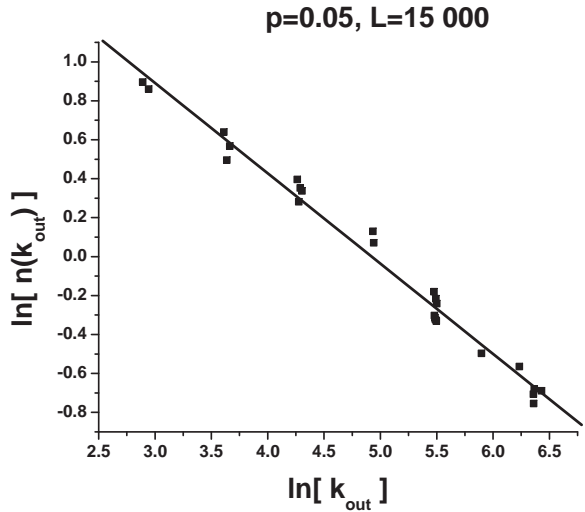}
\caption{The log-log fit to the envelope of the out-degree distribution for 
$p=0.05$, $L=15000$ (Fig.(\ref{zout2})).  The fit is to a slope of -0.46. 
} \label{fits2}
\ec
\end{figure}

\begin{figure}
\bc
\leavevmode
\includegraphics[width=8.5cm,angle=0]{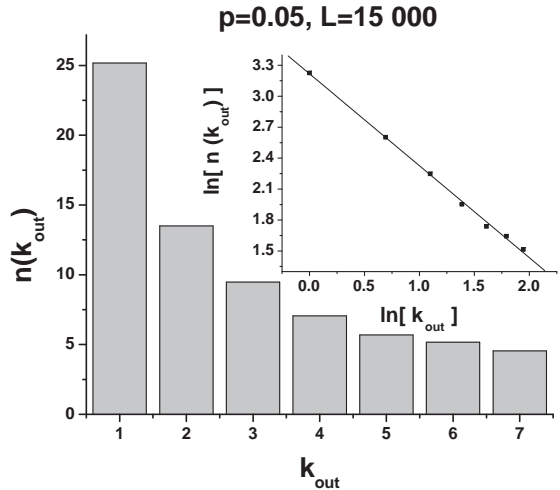}
\caption{The initial peak of the out-degree distribution for  $p=0.05$ and 
$L=15000$, and the double logarithmic graph of the same, showing incipient power 
law behaviour with the power $-0.89$. } \label{zout2begin}
\ec
\end{figure}

We display in Figs.(\ref{zout1},\ref{zout2}) the out-degree distribution
$n(k\ssout)$ for two sets of parameters, namely $L=5000,\;p=0.1$ and for
$L=15000\; p=0.05$, with, the number of genes fluctuating around $ \overline{N}=450 $
and $ \overline{N}=712$, respectively.  The data have been averaged over 
500 independent realizations.  The distributions, which are strongly log 
periodic,
with a power law envelope $n_m(k\ssout)\sim k\ssout ^{\gamma^\prime}$,
have the same characteristic behaviour for the two cases.  The fits to the
envelope of the peaks are shown in Figs.(\ref{fits1},\ref{fits2}).  We see
a marked break in the log-log fit for larger $p$, with a crossover from
$\gamma^\prime\simeq -1$ to $\gamma^\prime=-0.45 \pm 0.06$ at about $\ln 
(k\ssout) = 2.5$.
This crossover behavior is just off-scale in  Fig.(~\ref{fits2}), as 
can be seen in Fig.(~\ref{zout2begin}), with the incipient scaling with a 
power of $-1$ again extending out to $\ln(k\ssout) \simeq 2$.

It is also worthwhile mentioning that the integral over the peaks to the
rightmost of the out-degree distribution simply gives the network size
$N$, as it should, because these peaks come from the smallest genes of
unit length which are embedded in all the other genes.  Introducing a
larger lower cutoff to the gene size (i.e., a cutoff bigger than unity)  
would somewhat restrict the range of $k\ssout$, but also yield cleaner
scaling behaviour for large values of $k\ssout$ by effectively eliminating
these peaks.

It should be noticed that for $k\ssout \le 100$, the dips in the degree 
distribution are about evenly
spaced on the logarithmic scale, with a scale factor of $1/2$. We 
conjecture that each individual peak corresponds to a particular 
gene-length $\ell$. (This conjecture is borne out by comparing the integrals 
under the peaks with $n_\ell$ for $p=0.1, \; L=5000$ but is much more 
noisy for $p=0.05,\; L=15000$). 
Increasing $\ell$ by unity, exactly halves the leading contribution to the 
to $k\ssout$, from genes of length $\ell$. 

\begin{figure}
\bc
\leavevmode
\includegraphics[width=8.5cm,angle=0]{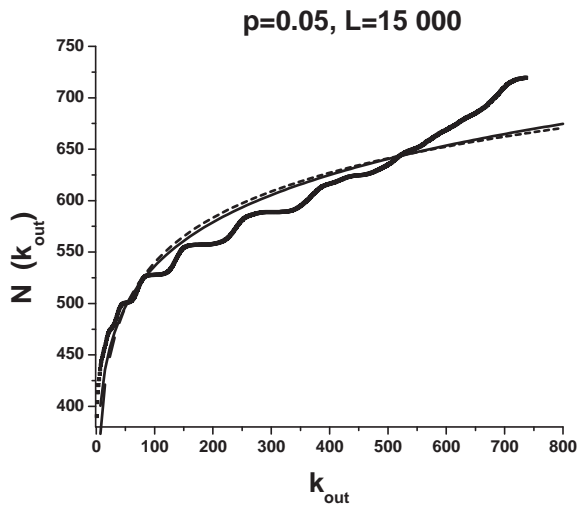}
\caption{The wavy line  is the cumulative out degree distribution for $p=0.05$, 
$L=15000$. The dashed line is a logarithmic fit, while the continuous one $\sim 
k\ssout ^{0.11}$.   } \label{cumul}
\ec
\end{figure}

To get  a better grip on the degree distribution, we numerically integrated the 
curves  in Figs.(\ref{zout1},\ref{zout2}).  The result is shown for $p=0.05$ in 
Fig.(\ref{cumul}).  We find that 
\be 
{\cal N}(k\ssout) \equiv \int^{k\ssout} n(z) dz \sim \ln(k\ssout)\;\;\;,
\ee
or a very small power, $\sim k\ssout ^{\gamma +1}$ where numerically 
$\gamma +1 < 0.1$, at least for $k\ssout$ that are not too large.  Thus we are 
led to believe that the overall out degree 
distribution scales like
\be
n(k\ssout) \sim k\ssout^{\gamma} f(k\ssout/\overline{N}) \;\;\;,
\ee
with $\gamma= -1.0 \pm 0.1$. The scaling function $f(x) \sim 
{\rm const.}$ 
for $x\ll 1$,  and is log-periodic for intermediate values of its 
argument. Scaling breaks down for $x\sim O(1)$.
Note that $\gamma$ is smaller in absolute value than that reported for 
real gene regulation 
networks~\cite{Sole,Maslov1,Maslov2}, thus evolution seems to have narrowed the 
distribution somewhat, restricting the number of genes that may be affected by 
any one gene.  

\begin{figure}
\bc
\leavevmode
\includegraphics[width=8.5cm,angle=0]{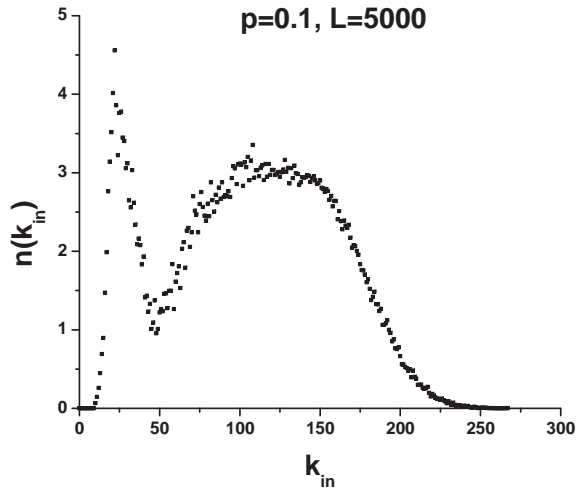}
\caption{The in-degree distribution for $p=0.1$ and $L=5000$.  The in-degree 
distribution is much narrower that the out-degree distribution and displays two 
peaks.  } \label{zin}
\ec
\end{figure}

\begin{figure}
\bc
\leavevmode
\includegraphics[width=8.5cm,angle=0]{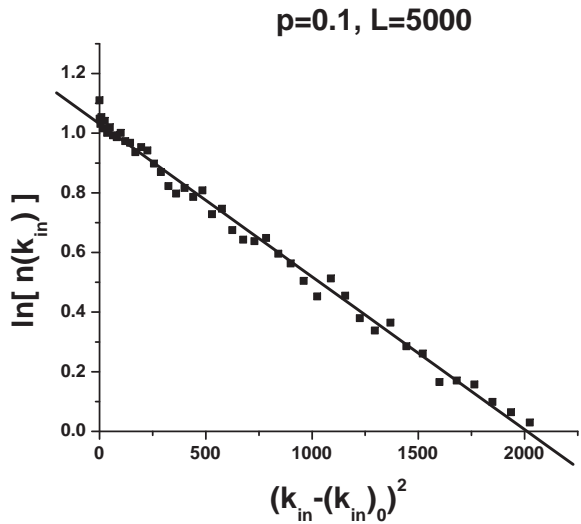}
\caption{The second peak of the in-degree distribution for $p=0.1$ and $L=5000$ 
can be fit to a Gaussian.} \label{Gauss}
\ec
\end{figure}

\begin{figure}
\bc
\leavevmode
\includegraphics[width=8.5cm,angle=0]{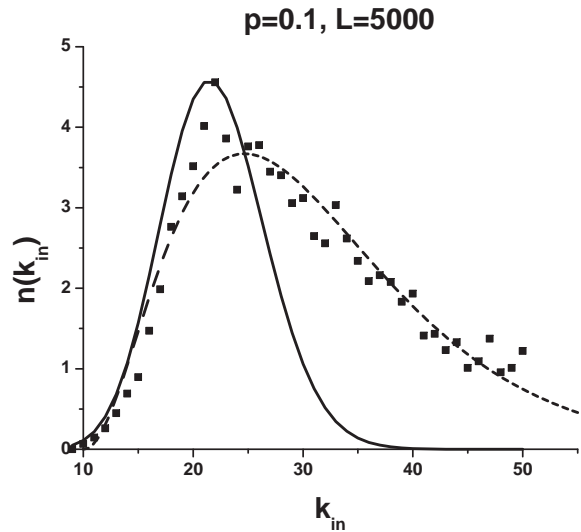}
\caption{The first peak of the in-degree distribution, with the tail of the 
Guassian distribution subtracted.  The fits are to a Poisson distribution 
(continuous line) $ \exp(-z) z^k/k!$ 
with ${ z=\overline k}$ computed from the data points, and the function $f(k) 
=0.09 (k-10)^{2.2} \exp [-0.15 (x-10)]$ (dashed line).} \label{1stpeak}
\ec
\end{figure}

The in-degree distribution, which we display in Fig.(\ref{zin}) shows two peaks.  
The second can be fit to a Gaussian, as shown in Fig.(\ref{Gauss}).  The first 
peak is more skewed than a Poissonian, but may be fitted reasonably well by a 
distribution of the form $f(x) \sim (x-x_0)^\delta \exp[-\xi (x-x_0)]$, 
where $\delta= 2.2$ and $\xi =0.15$.  (Fig.(\ref{1stpeak})).

\begin{figure}
\bc
\leavevmode
\includegraphics[width=8.5cm,angle=0]{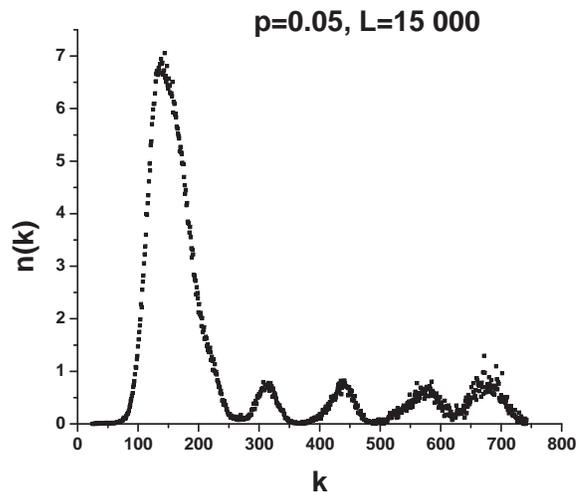}
\caption{The total degree distribution for $p=0.05,$ $L=15000$. } \label{all}
\ec
\end{figure}

The total degree distribution again displays a modulated structure, as can be 
seen in Fig.(\ref{all}).

\begin{figure}
\bc
\leavevmode
\includegraphics[width=8.5cm,angle=0]{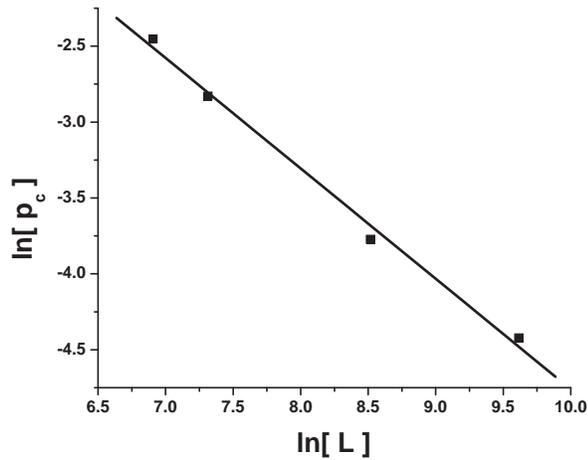}
\caption{Dependence of the ``percolation threshold'' on $L$.  The power 
law fit gives $p_c \sim L^{-3/4}$. } 
\label{pc}
\ec
\end{figure}

We have already discussed that the network will behave trivially either
for very small $N$ (all very large genes with very small probabilities for
interaction)  or for large $p$ (i.e., $p>1/2$) where most of the
chromosome will be occupied by non-coding partitions (the symbol 2), and
many null genes. We have determined the threshold value, $p_c$, below 
which the the genes are too few and too long, such 
that the single giant cluster breaks up into more than one component. The results are reported in Table I, for $L$ ranging from 15000 to 1000, and undirected edges. Although we have only a few points, the dependence (see Fig.~\ref{pc}) fits a power law, with 
$p_c \sim L^{-\alpha}$, with $\alpha \simeq 3/4$, within the range explored.
In the  limit of $L\to \infty$, $p_c \to 0$, but less fast than $1/L$.

\begin{table}
{\bf Table I} 
\vskip 1cm
\begin{ruledtabular}
\begin{tabular}{||c|c|c||}
\hline
 $L $ &  $\langle N \rangle$  &  $p_c$ \\ \colrule
15000     &  177    & 0.012       \\
5000    & 112     &  0.023      \\
1500 &  83  & 0.059  \\
1000 & 54      & 0.086\\
\colrule 
\end{tabular}
\end{ruledtabular}
\caption{ The ``percolation threshold'' for strings of different length $L$. The
average number of non-null genes at the percolation threshold is also
reported.} \end{table}

On the other hand our expectations that there is a reasonably
large interval of $p$'s ($p_c < p <1/2$) where the results do not depend 
quantitatively 
on the precise value of $p$ are borne out, as can be seen from  our 
calculations  for 
the minimum average path length for undirected edges.

We argued in section II that as long as there is a single giant cluster, 
the directed minimum path length is identically 1.  For undirected paths 
this is no longer true and we must determine $\langle l_{\rm min} \rangle$ 
numerically.  Nevertheless, we can show that $l_{\rm min} \le 4$.
Note that for $p>p_c$, there will be an abundance of short genes of unit
length, which will contain the symbol 0 or 1 with equal probability. These
will have outgoing edges to all the typical genes which have an admixture
of $0$s and $1$s, so that most such genes will be certainly linked by
paths of length at most equal to 2.  The ``worst case'' is that of
atypical cases of all 1's all or all 0's which require at least one
intermediate typical gene to link up ($11111 - 1 - 011100 - 0 - 00000$),
which gives ${\rm max}( l_{\rm min}) =4$.  We have calculated $\langle 
l_{\rm min}
\rangle$ for fixed $p$ and different $L$. 
 If we consider undirected edges
and find that it at most depends very weakly on $N$, the number of 
vertices.  We
find, e.g., for $L$ ranging from 1650 to 15000, and $p=0.05$ that $\langle
l_{\rm min} \rangle$ ranges only from 1.673 to 1.699. (Table II)

\begin{table}
{\bf Table II} 
\vskip 1cm
\begin{ruledtabular}
\begin{tabular}{||c|c|c||}
\hline
L     & $\langle N\rangle$ & $\langle l_{\rm min}\rangle$  \\ \colrule
15000     & 717       & 1.669   \\
10000 &    479     & 1.670      \\
5000 & 239     & 1.670  \\
2500     & 120      & 1.670\\
2000  & 95      & 1.669  \\ 
1750  & 83 & 1.672  \\
1650  & 79   & 1.673 \\
\colrule 
\end{tabular}
\end{ruledtabular}
\caption{The average minimum 
path length for undirected edges, with $p=0.05$, for different $L$.}
\end{table}

Fixing the length $L$ and varying $p$ (Table III) shows again  that as 
long as $\langle l_{\rm min}\rangle$ is defined, i.e., just above the 
``percolation threshold'' it is already very close to the value it will have for larger $p$.

\begin{table}
{\bf Table III} 
\vskip 1cm
\begin{ruledtabular}
\begin{tabular}{||c|c|c|c||}
\hline
 $p$  & $L $    & $\langle N\rangle$ & $\langle l_{\rm min}\rangle$  \\ 
\colrule
0.013 & 15000    & 194     & 1.858   \\
0.014 & 15000    & 209     & 1.851      \\
0.016 & 15000    & 237     & 1.84  \\
0.018 &15000     & 267     & 1.82\\
0.02  & 15000    & 295     & 1.81 \\ 
0.05  & 15000   & 717     &  1.669 \\
0.1   & 10000   & 452     & 1.543 \\
0.48  & 1500    & 374     & 1.360 \\
\colrule 
\end{tabular}
\end{ruledtabular}
\caption{The average minimum path length $\langle l_{\rm min}\rangle$ for 
undirected edges,  for different $p$ and $L$. Note that $\langle 
l_{\rm min}\rangle$ varies very little over the entire range from $p 
\simeq p_c $ to $p\simeq  1/2 $. } \end{table}

%Kumelenme katsayilari soyle:
%<Cout>=0.034035
%<Cin>=0.648129
%<C>=0.534313    

The clustering coefficients are markedly different for the in- and out-degree 
distributions, as well as the total connectivity.  We find, for $p=0.05$ and 
$L=15000$,  $<C\ssout> = 0.034$, $<C\ssin >= 0.648$ and $<C>=0.534$.  We have 
verified in the last case that this clustering coefficient is reproduced by 
computing the average total degree per node and deviding by the total number of 
nodes, i.e., $<C>\simeq <k> / N $. Thus, the total connectivity behaves 
very 
much like in classical random graphs.

\section{Analytical results on toy model}

In this section we would like to present a number of the results which may
be obtained from a hierarchical model.
In particular, we would like to exploit the transitivity property, to 
compute certain properties of a toy version of our original model.

Let us consider a tree network, with a branching ratio $b$. Take  all 
connections emanating from nodes at the $m$th generation to the ones at 
the $m+1$st generation, to be directed.  If one applies the transitivity 
rule, then this automatically generates further connections that link 
directly a node at the $m$th generation to all the nodes below it on the 
same branch of the tree.  Then, the out degree (we will drop the index 
``out" from now on) of a node at the $m$th level will be 
\be 
k(m) = \sum_{m^\prime=0}^{M-m-1} b^{m^\prime} = { b^{M-m} - 1 \over 
b-1}\;\;\;,
\ee
where $M$ is the total number of levels and $k(m) \sim b^{M-m}$ for large 
$M-m$, which we may safely assume.  The frequency of the nodes at the 
$m$th level is  $b^{m}$.  This gives, for the out-degree 
distribution,
\be
n(k) \sim {\rm const.} \left( { 1 \over k} \right) \;\;\;.
\ee

We now introduce redundancies within the same level. Let a fraction $r$ of 
the 
$b$ downstream neighbors emanating from any node be identical to each other, 
with $r=a^m$, $ b^{-1} < a < 1 $ being some constant.
The nodes that are identical within any given level are
connected to each other by two-way bonds within the sub-branch where they are 
located.  I will call this subset of nodes the clones at any given level. This 
means that the out-degree 
$k$ of any one of these nodes is now $k^\prime = (a^m b) b^{M-m}$, since 
each 
interconnected node inherits all the downstream neighbors of the  $a^m b$ clones 
to which it is connected.  The interconnections between nodes further downstream 
do not introduce any further change in $k^\prime(m)$, since all the nodes that 
they can connect are already connected to the nodes at level $m$. The number of 
clone-nodes (not all identical to each other, but only in groups of $a^m 
b$) is 
clearly $a^m b^m$. Then, the out-degree distribution of the nodes belonging to 
the clone-sets becomes,
\be 
n(k^\prime) = b^{\gamma -1} \left( {k^\prime \over N} \right) 
^{\gamma}  \;\;\;
\ee
where $N=b^M$ and with 
\be
\gamma ={ \ln a + \ln b \over \ln a - \ln b} \;\;\;.
\ee
Note that under our assumptions above, $-1 < \gamma < 0 $.  
For large enough $k$ this then becomes the {\it leading} contribution to the out 
degree distribution, rather than $\gamma = -1$. 

\section{Conclusions and Discussion}

The model we have presented for the scaling behaviour of gene interaction, 
more specifically RNA interference, 
turns out to be a rich model which is worthwhile to investigate in its 
own right, as well as 
providing a highly structured starting point on which further evolutionary 
pressures could act.

We have checked that the statistical properties of the network are robust under random point mutations.  Thus, the scale free random network may be seen as the attrator for a process starting from, say, a sequence of uniform genes, subjected to random point mutations over a very long time period.  The scale free steady state can be considered as the outcome of a process of self-organization under random perturbations.

In conclusion it is worthwhile to note that in RNA interference, longer 
RNA chains are cut up by the ``Dicer" enzyme into smaller segments, 
siRNA~\cite{Hannon}, which 
can be conveniently matched with complementary sequences in a larger 
number of mRNAs.  This trick would change the power of the out-degree 
distribution from that found here. 

{\bf Acknowledgements}
\nopagebreak
\smallskip
One of us (AE) would like to gratefully acknowledge partial support from
the Turkish Academy of Sciences.  It is a pleasure to thank B. Bollobas 
and J.-P. Eckmann for useful conversations.

\appendix

{\bf APPENDIX}
\nopagebreak
\smallskip
We would like to compute the probabilities $P(\ell, \ell^\prime)$ in
Eq.(\ref{P}). For successive $\nu=\ell^\prime-\ell>0$, it is convenient to
define subsequences of $G_j$ of length $\ell_i$ with a shift $\lambda$,
$0\le \lambda \le \nu $, such that
$g_{ij}^{(\lambda)} \equiv \{x_{j,1+\lambda}, x_{j,2+\lambda}, \ldots
x_{j,\ell_i+\lambda}\}$.
Then, we can obtain $P(\ell, \ell^\prime)$ in terms of joint probabilities like 
${\cal P}(G_i\neq g_{ij}^{(0)}, G_i = g_{ij}^{(1)})$, etc.  Thus, 
\begin{eqnarray*}
P(\ell, \ell) &&= {\cal P}(G_i=g_{ij}^{(0)})\cr
P(\ell, \ell+1) &&= {\cal P}(G_i=g_{ij}^{(0)}) + {\cal P}( G_i\neq 
g_{ij}^{(0)}, G_i=g_{ij}^{(1)})\cr
&\ldots& \cr
P(\ell, \ell+\nu)&&= {\cal P}(G_i=g_{ij}^{(0)}) +\ldots \cr
&+& {\cal P}( G_i\neq 
g_{ij}^{(0)}, G_i \neq g_{ij}^{(1)}, \ldots, G_i = G_{ij}^\lambda ) \cr 
&+& \ldots \cr
&+&{\cal P}( G_i\neq g_{ij}^{(0)}, G_i \neq g_{ij}^{(1)}, 
\ldots, 
G_i = g_{ij}^{(\nu)} ) 
\end{eqnarray*}

Let us  display ${\cal P}(  G_i\neq g_{ij}^{(0)}, G_i=g_{ij}^{(1)})$ for 
greater 
clarity.  We have
\begin{eqnarray*} 
\lefteqn{{\cal P}( G_i\neq g_{ij}^{(0)}, G_i=g_{ij}^{(1)}) 
= }\cr
&& \langle 
\left(\sum_{\mu = 1}^\ell 
 \prod_{m=1}^{\mu-1}   \delta_{x_{i,m} x_{j,m}} \right)  
\left(1-\delta_{x_{i,\mu},x_{j,\mu}}\right)
\prod_{n=1}^{\ell_i} \delta_{x_{i,n} x_{j,n+1}} \rangle \cr
&&= 2^{-\ell_i}\left[1-2^{-\ell_i}\right]\;\;.
\end{eqnarray*}

In particular we find
\begin{eqnarray*}
P(\ell, \ell+1) &=& 2^{-\ell}\left[1+ 1-2^{-\ell}\right]\cr
&=& 2^{-\ell}\left[2 -2^{-\ell}\right]\cr
P(\ell, \ell+2)&=&2^{-\ell} \left[3-2^{-\ell+1} - {1 \over 3} 
\left(1- 2^{-2\ell}\right) \right]\cr
\end{eqnarray*}
Note that the probabilities ${\cal P}$ do not depend upon the particular
gene, except through the lengths of the sequences, as the genes do not
overlap, and each of the symbols are independently and identically
distributed, with equal probability, over 0 and 1.

\end{document}